\documentclass{aa}  
\usepackage{graphicx}
\usepackage{txfonts}
\newcommand{\kms}{\ensuremath{\mathrm{km\,s}^{-1}}}
\newcommand{\vsi}{\ensuremath{v_\mathrm{e} \sin i}}

\newcommand{\te}{\ensuremath{T_{\mathrm{e}}}}
\newcommand{\logg}{\ensuremath{\log g}}

\begin{document}
   \title{Atmospheric velocity fields in tepid main sequence
     stars\thanks{Based in part on observations obtained at the
       Canada-France-Hawaii Telescope (CFHT) which is operated by the
       National Research Council of Canada, the Institut National des
       Sciences de l'Univers of the Centre National de la Recherche
       Scientifique of France, and the University of
       Hawaii.}\fnmsep\thanks{Based in part on observations made at
       Observatoire de Haute Provence (CNRS), France.}}


   \author{J. D. Landstreet 
          \inst{1,2}
          \and
          F. Kupka  
          \inst{3,6}
          \and
          H. A. Ford  
          \inst{2,4,5}
          \and
          T. Officer 
          \inst{2}
          \and
          T. A. A. Sigut  
          \inst{2}
          \and
          J. Silaj  
          \inst{2}
          \and
          S. Strasser 
          \inst{2}
          \and
          A. Townshend 
          \inst{2}
          }

   \offprints{J. D. Landstreet}

   \institute{Armagh Observatory, College Hill, Armagh BT61 9DG,
             Northern Ireland
         \and
             Department of Physics \& Astronomy, University of Western
             Ontario, London, ON N6A 3K7, Canada 
         \and
             Max-Planck-Institut f\"ur Astrophysik,
             Karl-Schwarzschild-Str. 1, 85748 Garching, Germany 
         \and
             Centre for Astrophysics and Supercomputing, Swinburne 
             University of Technology, Hawthorn, Victoria 3122, Australia.
         \and
             Australia Telescope National Facility, CSIRO, Epping, NSW 1710,
             Australia.
         \and
             Observatoire de Paris, LESIA, CNRS UMR 8109, F-92195, Meudon,
             France
}

   \date{Received September 1, 2009; accepted August 1, 2019}

  \abstract
%
  {The line profiles of the stars with \vsi\ below a few \kms\ can
    reveal direct signatures of local velocity fields such as
    convection in stellar atmospheres. This effect is well established
    in cool main sequence stars, and has been detected and studied in
    three A stars.  }
%
%
  {This paper reports observations of main sequence B, A and F stars
    (1) to identify additional stars with sufficiently low values of
    \vsi\ to search for spectral line profile signatures of local
    velocity fields, and (2) to explore how the signatures of the
    local velocity fields in the atmosphere depend on stellar
    parameters such as effective temperature and peculiarity type.  }
%
%
  { We have carried out a spectroscopic survey of B and A stars of low
    \vsi\ at high resolution. Comparison of model spectra with those
    observed allows us to detect signatures of the local velocity
    fields such as asymmetric excess line wing absorption, best-fit
    \vsi\ parameter values that are found to be larger for strong
    lines than for weak lines, and discrepancies between observed and
    modelled line profile shapes. }
%
%
  { Symptoms of local atmospheric velocity fields are always detected
    through a non-zero microturbulence parameter for main sequence
    stars having \te\ below about 10000~K, but not for hotter
    stars. Direct line profile tracers of the atmospheric velocity
    field are found in six very sharp-lined stars in addition to
    the three reported earlier. Direct signatures of velocity fields
    are found to occur in A stars with and without the Am chemical
    peculiarities, although the amplitude of the effects seems larger
    in Am stars. Velocity fields are also directly detected in
    spectral line profiles of A and early F supergiants, but without
    significant line asymmetries.}
%
%
  { We confirm that several atmospheric velocity field signatures,
    particularly excess line wing absorption which is stronger in the
    blue line wing than in the red, are detectable in the spectral
    lines of main sequence A stars of sufficiently low \vsi. We triple
    the sample of A stars known to show these effects, which are found
    both in Am and normal A stars. We argue that the observed line
    distortions are probably due to convective motions reaching the
    atmosphere.  These data still have not been satisfactorily
    explained by models of atmospheric convection, including numerical
    simulations.}


   \maketitle
%

\section{Introduction}

It has been known for many years that the spectral line profiles of
sharp-line cool main sequence stars similar to the Sun are
asymmetric. When the observed line profiles of such stars are compared
to symmetric theoretical profiles, it is apparent that the observed
profiles in general have a slightly deeper long-wavelength wing
compared to the short-wavelength wing. This asymmetry is often
described by means of a line bisector (a curve bisecting horizontally
the observed absorption profile). Such line bisectors in general have
the approximate form of a left parenthesis ``(``. The line bisectors
of these stars are usually described as ``C''-shaped.

This effect is readily visible in a large number of lower main
sequence stars because many such stars have the very low projected
rotational velocity values ($\vsi \leq 5 - 8$~\kms) required for the
effect to be detectable at reasonable resolving power ($\sim 10^5$)
and signal-to-noise ratio ($SNR \sim 200-500$). A large survey of
bisector curvature among cool stars (spectral type mid-F and later) of
luminosity classes II--V is reported by Gray (\cite{Gray05}).

Such line asymmetry is generally interpreted as evidence for the
presence of an asymmetric convective velocity field in the stellar
photosphere, with most of the atmosphere rising slowly (and thus
slightly blue-shifted), while over a small fraction of the photosphere
there are relatively rapid downdrafts (thus slightly red-shifted). The
observed combination of the spectral lines from these two regions
leads to integrated line profiles dominated by the slowly rising
material but showing weak red-shifted absorption wings from the
downdrafts, and line bisectors with a roughly ``C'' shaped profile. 

This qualitative picture is strongly supported by 3D hydrodynamic
numerical simulations of the convection in the outer layers of cool
stars. Such calculations exhibit the expected behaviour of large
slowly rising gas flows that are balanced by more rapid downflows over
more limited regions, and in fact the predicted line profiles are in
very good agreement with observed profiles (e.g. Dravins \& Nordlund
\cite{DravNord90}; Asplund et al. \cite{Aspletal00}; Allende
Prieto, Asplund, L\'opez \& Lambert \cite{Prieetal02}; Steffen et
al. \cite{sfl06}; Ludwig \& Steffen \cite{LudwStef08}).

Until recently little was known about the situation farther up the
main sequence, mainly because above an effective temperature of 
$\te \sim 7000$~K, most main sequence stars have \vsi\ values on the order of
$10^2$~\kms, which is much larger than the $\sim 1 - 10$~\kms\ values
of cool stars. This rotational broadening completely overwhelms any
bisector curvature. In a study of bisector curvature of cooler stars,
Gray \& Nagel (\cite{gn89}) noted that one Am (A8V) star, HD~3883,
exhibits ``reversed'' bisector curvature; that is, that the depressed
line wing is on the short-wavelength side of the line rather than on
the long-wavelength side. They also found such reversed bisectors in
several giants and supergiants lying on the hot side of a line in the
HR diagram running from about F0 on the main sequence to G1 for type
Ib supergiants. Gray \& Nagel argued that their data are evidence for
a ``granulation boundary'' in the HR diagram, with convective motions
on the hot side of this boundary being structurally different from
those on the cool side.

This picture was extended by Landstreet (\cite{jdl98}), who searched
for A and late B stars (``tepid'' stars) of very low \vsi, and then
carefully modelled the line profiles of such stars. He discovered two
Am stars and one normal A star that show clear reversed bisectors, but
found that this effect is absent from late B stars. Landstreet further
established that reversed bisectors appear to correlate well with the
presence of measurably non-zero microturbulence.  Consistent with the
results of testing theoretical model atmospheres for convective
instability using the Schwarzschild criterion, this microturbulence
dies out at about A0. There is thus a {\em second} ``granulation
boundary'' in the HR diagram, above which granulation (and apparent
photospheric convection) is absent.

Landstreet also showed that the line profiles of strong spectral lines
in the late A (Am) stars are very different from those predicted by
line synthesis using the simple model of symmetric Gaussian
microturbulence, and argued that such line profiles carry a
considerable amount of information about stellar photospheric velocity
fields.

Recently the first 3D numerical simulations of convection in A stars
have been carried out by Steffen, Freytag \& Ludwig (\cite{sfl06})
(see also Freytag \& Steffen \cite{fs04} and the earlier simulations
in 2D by Freytag, Steffen \& Ludwig \cite{fsl96}). These 3D numerical
simulations show line profile distortions with approximately the right
amplitude (bisector spans of 1 -- 2 \kms\ compared to observed values
about twice that large for $\te \sim 8000$~K), but the depressed line
wing is the long-wavelength one, as in the solar case. There is
currently no theoretical explanation of the observed depressed blue
line wings in A stars, or indeed anywhere on the hot side of the
granulation boundary. (Note that Trampedach [\cite{Tram04}] has also
carried out computations of convection in the outer envelope of an A9
star, without computing line profiles, but at this temperature it is
quite possible that the star lies on the low temperature side of the
granulation boundary.)

At present, empirical information about velocity fields in main
sequence A and B stars is very limited. For most such stars, the only
information available is what can be deduced from the microturbulence
velocity parameter. Useful local line profile information is available
for only a handful of A and late B stars. Furthermore, the best
available theoretical models do not predict the observed line
asymmetries.  In these circumstances, it would be very useful to have
more stars in which line asymmetries are visible directly, both to
study how convective phenomena vary with stellar parameters such as
mass, \te\ and spectral peculiarity, and to have a larger sample of
observed profiles with which to compare the new generation of
numerical simulations now being produced.

Consequently, we have carried out a survey of A and B stars (mostly on
or near the main sequence) with the goal of identifying more stars
with sufficiently small \vsi\ values that the atmospheric velocity
field affects observed line profiles. As stars of really small
\vsi\ are found, we have modelled them to determine what symptoms of
velocity fields are contained in the profiles. We have also determined
a number of parameters, including  effective temperature \te,
 gravity \logg, abundances of a few elements, accurate values of
\vsi, and (where possible) the microturbulence parameter $\xi$, for
the full dataset presented here. The goal of this survey is to provide
a significantly larger, and reasonably homogeneous, body of
observations that may be used to expand our knowledge of velocity
fields in the atmospheres of A and late B stars.

\section{Observations and data reduction}

The principal difficulty in detecting and studying the effects of
photospheric convection (or other local velocity fields) on line
profiles in A and B stars on or near the main sequence is that most
have rotation velocities of the order of $10^2 - 3\,10^2$ \kms. In
this situation, most velocity effects on the local line profiles are
obliterated by the rotational broadening. The single effect that
persists even at high \vsi\ is the requirement for non-zero
microturbulence (characterised by the microturbulence parameter
$\xi$), needed to force abundance determinations using both weak and
strong lines to yield the same abundance values. This parameter is
widely thought to be a proxy for the presence of photospheric
convection; Landstreet (\cite{jdl98}) has shown that among main
sequence A and B stars, $\xi$ goes to zero at approximately the same
effective temperature (around 10500~K) where the Schwarzschild
instability criterion applied to Kurucz (e.g. \cite{Kuru79}) model
atmospheres shows that the atmospheres become marginally or fully
stable against convection.  (Note that convection becomes {\em
  energetically} unimportant in the construction of such model
atmospheres, because it is so inefficient that it carries negligible
heat, at a substantially lower \te\ value, $\sim 8250$~K, and is
usually neglected above this temperature.) The microturbulence
parameter is essentially the only kind of observational information
available at present for the study of atmospheric convection in most
main sequence A and B stars.

However, a few tepid stars do have small enough values of \vsi, less
than a few \kms, for the effects of the velocity field of surface
convection to be directly detectable in the observed line profiles.
Three A stars (two Am stars and one normal one) in which the effects
of local velocity fields are visible in the lines have been identified
and studied by Landstreet (\cite{jdl98}); three other late B or early
A stars (two HgMn stars and one normal star) were also found, in which
neither microturbulence nor line profile distortions (relative to
simple models) reveal any hint of photospheric convection. These data
suggest that convection becomes detectable in line profiles of main
sequence stars below $T_e \sim 10000$~K, but clearly with a sample so
small, conclusions must be rather provisional. An essential first step
to drawing more secure conclusions from observations is to obtain a
larger sample of stars in which velocity fields can be studied
directly in the line profiles.

In order to enlarge the sample of main sequence A and B stars of very
low \vsi\ which may be studied, we have carried out a substantial
observational survey. In spite of the large number of \vsi\ values for
A and B main sequence stars in the current literature (see for example
Abt \& Morrell \cite{AbtMorr95}; Levato et al.  \cite{Levaetal96};
Wolff \& Simon \cite{WolfSimo97}; Abt, Levato \& Grosso
\cite{Abtetal02}; and references therein), a survey was necessary
because most of the available observations of rotational velocities of
early-type stars have been carried out with a spectral resolving power
$R$ of the order of 1--3\,$10^4$. This resolving power is insufficient
to determine the value of \vsi\ if it is less than 10--20 \kms. Since
Landstreet (\cite{jdl98}) found that direct detection of velocity
fields in line profiles requires \vsi\ to be less than about 5 to
10~\kms, new observations were required to identify stars satisfying
this criterion.

To have adequate resolving power to identify stars with \vsi\ values
below a few \kms, we have used the high-resolution spectrograph
Aur\'elie on the 1.52-m telescope at the Observatoire de Haute
Provence (OHP) in France, and Gecko at the 3.6-m Canada-France-Hawaii
telescope (CFHT).  Both have $R \approx 1.2\,10^5$. Spectra obtained
with these instruments can measure \vsi\ values down to about 2
\kms. This resolving power is sufficient to (barely) resolve the
thermal line width of Fe peak elements, which have mean thermal
velocities of about 1.7 \kms, and thus FWHM line widths of about 3
\kms. Both spectrographs are cross-dispersed echelle instruments, but
in each, only one order is observed, so that observations record only
a single wavelength window, typically 30 to 80~\AA\ in length,
depending on the detector used.

Observing missions were carried out at OHP in 1993, 1994 and 1995. For
all missions, a small window covering 4615 -- 4645~\AA\ was used,
isolated by a custom interference filter. Observations were carried
out at CFHT in 1993 and twice in 1995, using the same observing
window, and also a window covering 5150 -- 5185~\AA. The data from
these runs have been discussed by Landstreet (\cite{jdl98}). Further
missions were carried out in 2000 and 2001 at CFHT, both to look for
more very sharp-line A and B stars, and to increase the available data
for stars already identified as particularly interesting. For these
runs, a larger CCD was available, and windows at 4525 -- 4590~\AA\ (in
2000) and 4535 -- 4600~\AA\ (in 2001) were isolated by a grism. 

Candidate stars for observation were largely chosen from stars which
are reported to have values of \vsi\ below about 15 -- 20 \kms\ in
previous surveys of rotational velocity (see references
above). Although these surveys lack the resolving power to identify
directly the stars of most interest here, they do constitute a first
sorting which select the few percent of all early type main sequence
stars that may be really sharp-lined; without these surveys to guide
us, the required observing programme would have been more than ten
times larger. 

Most of the stars observed are B, A, and early F main sequence stars,
although a few bright giants and supergiants were also observed.  Examples
from the Am and HgMn peculiarity groups were included in the survey.
All the stars observed are brighter than about $V \approx 7$, since
this is the effective limiting magnitude of most of the previous \vsi\
surveys available.  The spectra obtained typically have
signal-to-noise ratio (SNR) of the order of 200 in the continuum.
Stars that are found to have very small \vsi\ values are likely to be
members of close binary systems, which may not have been previously
noticed, and so the stars of greatest interest to us, with \vsi\ below
8 -- 10 \kms, were mostly observed more than once to test for the
presence of a secondary in the observable spectrum.

Spectra were reduced using standard IRAF tasks: CCD frames were
corrected for bias and bad pixels and extracted to 1D; the 1D
spectra were divided by similarly extracted flat field exposures; a
polynomial was fit to continuum points to normalise the continuum to
1.0; the spectra were wavelength calibrated, using typically 30 ThAr
lines which are fit to an RMS accuracy of about $3\,10^{-4}$~\AA;
finally, the spectra were corrected to a heliocentric reference frame.

\begin{table*}
\begin{center} 
\caption{\label{all-good-stars} Sharp-line programme stars} 
\begin{tabular}{llllllllllll} 
\hline\hline 
 HD & other & spectral & binarity & \te & log $g$ & \vsi & $\xi \pm \sigma$ & log $n_{\rm Si}/n_{\rm H}$ & log $n_{\rm Cr}/n_{\rm H}$  & log $n_{\rm Fe}/n_{\rm H}$  \\ 
    &       & type     &      & (K)  & (cgs) & (\kms) & (\kms) \\ 
\hline
2421       & HR 104     & A2Vs       & SB2        &  10000 & 4.0   &  3.5 $\pm$    1 &  1.0 $\pm$  0.4 & -4.62 $\pm$ 0.14 & -6.25 $\pm$ 0.12 & -4.55 $\pm$ 0.14  \\ 
27295      & 53 Tau     & B9 HgMn    & SB1        &  11900 & 4.25  &  4.9 $\pm$  0.3 &  0.4 $\pm$  0.4 & -4.60 $\pm$ 0.11 & -5.85 $\pm$ 0.11 & -5.32 $\pm$ 0.14  \\ 
27962      & 68 Tau     & A2IV m     & SB1        &   8975 & 4.0   &   11 $\pm$  0.5 &  2.0 $\pm$  0.3 &              & -5.75 $\pm$ 0.14 & -4.20 $\pm$ 0.14  \\ 
46300      & 13 Mon     & A0Ib       &            &   9700 & 2.1   &   10 $\pm$    1 &    4 $\pm$  0.5 &              & -6.80 $\pm$ 0.14 & -4.90 $\pm$ 0.14  \\ 
47105      & $\gamma$ Gem & A0IV       & SB1        &   9150 & 3.5   &   11 $\pm$  0.4 &  1.2 $\pm$  0.4 &              & -6.25 $\pm$ 0.12 & -4.60 $\pm$ 0.12  \\ 
48915      & $\alpha$ CMa & A1V m      & SB1 VB     &   9950 & 4.3   & 16.5 $\pm$    1 &  2.1 $\pm$  0.3 &              & -5.80 $\pm$ 0.18 & -4.27 $\pm$ 0.12  \\ 
61421      & $\alpha$ CMi & F5IV-V     & SB1 VB     &   6500 & 3.95  &    7 $\pm$    1 &  2.2 $\pm$  0.3 &              & -6.52 $\pm$ 0.14 & -4.78 $\pm$ 0.14  \\ 
72660      & HR 3383    & A1V        &            &   9650 & 4.05  &  5.0 $\pm$  0.5 &  2.3 $\pm$  0.3 & -4.20 $\pm$ 0.11 & -6.09 $\pm$ 0.10 & -4.35 $\pm$ 0.12  \\ 
73666      & 40 Cnc     & A1V        &            &   9300 & 3.8   & 10.0 $\pm$    1 &  2.7 $\pm$  0.3 &              & -6.15 $\pm$ 0.14 & -4.55 $\pm$ 0.14  \\ 
78316      & $\kappa$ Cnc & B8 HgMn    & SB1        &  13400 & 3.9   &  6.8 $\pm$  0.5 &  0.5 $\pm$  0.5 & -4.10 $\pm$ 0.11 & -6.16 $\pm$ 0.18 & -4.47 $\pm$ 0.12  \\ 
82328      & $\theta$ UMa & F6IV       & SB1        &   6360 & 4.1   &  9.0 $\pm$    1 &  2.2 $\pm$  0.2 &              & -6.62 $\pm$ 0.14 & -4.95 $\pm$ 0.14  \\ 
87737      & $\eta$ Leo & A0Ib       &            &   9700 & 2.0   &   11 $\pm$    1 &    5 $\pm$  0.5 &              & -6.65 $\pm$ 0.14 & -4.80 $\pm$ 0.14  \\ 
103578     & 95 Leo     & A3V        & SB2        &   8300 & 4.0   &    3 $\pm$    1 &  2.0 $\pm$  0.2 & -4.36 $\pm$ 0.18 & -6.73 $\pm$ 0.14 & -4.98 $\pm$ 0.12  \\ 
107168     & 8 Com      & A8 m       &            &   8150 & 3.8   &   13 $\pm$    1 &  3.0 $\pm$  0.5 & -4.45 $\pm$ 0.14 & -5.75 $\pm$ 0.14 & -4.15 $\pm$ 0.14  \\ 
108642     & HR 4750    & A2 m       & SB1        &   8100 & 4.1   &    2 $\pm$    2 &  4.0 $\pm$  0.5 & -4.57 $\pm$ 0.19 & -6.30 $\pm$ 0.14 & -4.55 $\pm$ 0.14  \\ 
109307     & 22 Com     & A4V m      &            &   8400 & 4.1   & 14.5 $\pm$  1.5 &  3.2 $\pm$  0.3 &              & -6.50 $\pm$ 0.14 & -4.63 $\pm$ 0.14  \\ 
114330     & $\theta$ Vir & A1Vs       &  VB2       &   9250 & 3.4   &    4 $\pm$    1 &  1.3 $\pm$  0.2 &              & -6.40 $\pm$ 0.14 & -4.80 $\pm$ 0.14  \\ 
120709     & 3 CenA     & B5 HgMn    &  VB        &  17000 & 3.7   &    2 $\pm$    2 &  0.8 $\pm$  0.8 & -4.10 $\pm$ 0.18 & -7.10 $\pm$ 0.31 & -4.20 $\pm$ 0.18  \\ 
128167     & $\sigma$ Boo & F2V        &            &   6750 & 4.6   &  9.7 $\pm$  0.4 &  1.6 $\pm$  0.3 &              & -6.65 $\pm$ 0.18 & -4.90 $\pm$ 0.14  \\ 
132955     & HR 5595    & B3V        &  VB?       &  16400 & 4.4   &    7 $\pm$    1 &   nd $\pm$   nd & -4.25 $\pm$ 0.11 & -6.63 $\pm$ 0.31 & -4.90 $\pm$ 0.18  \\ 
142860     & $\gamma$ Ser & F6IV       &            &   6250 & 4.5   & 11.5 $\pm$    1 &  1.5 $\pm$  0.3 &              & -6.55 $\pm$ 0.18 & -4.80 $\pm$ 0.14  \\ 
147084     & o Sco      & A4II-III   &            &   7850 & 2.0   &    5 $\pm$  0.5 & 2.75 $\pm$  0.3 &              & -6.60 $\pm$ 0.14 & -4.80 $\pm$ 0.14  \\ 
149121     & 28 Her     & B9.5 HgMn  &            &  11000 & 3.9   &  9.2 $\pm$  0.5 &  0.5 $\pm$  0.5 &              & -6.16 $\pm$ 0.11 & -4.48 $\pm$ 0.11  \\ 
157486     & HR 6470    & A0V        & SB2        &   9150 & 3.65  &  1.5 $\pm$    1 &  1.9 $\pm$  0.2 &              & -6.58 $\pm$ 0.14 & -4.78 $\pm$ 0.14  \\ 
160762     & $\iota$ Her & B3IV       & SB1        &  17000 & 3.9   &  6.5 $\pm$  0.5 &    1 $\pm$    1 & -3.95 $\pm$ 0.14 & -6.65 $\pm$ 0.18 & -5.00 $\pm$ 0.18  \\ 
170580     & HR 6941    & B2V        &  VB        &  18000 & 4.0   &  5.5 $\pm$    1 &   nd $\pm$   nd & -4.00 $\pm$ 0.14 &              & -4.90 $\pm$ 0.14  \\ 
172910     & HR 7029    & B3V        &  VB?       &  18800 & 4.4   &   10 $\pm$    2 &   nd $\pm$   nd & -4.02 $\pm$ 0.14 & -6.72 $\pm$ 0.18 & -4.90 $\pm$ 0.31  \\ 
174179     & HR 7081    & B3IV p     &            &  17400 & 3.9   &  1.5 $\pm$  1.5 &    1 $\pm$    1 & -4.45 $\pm$ 0.18 & -6.58 $\pm$ 0.18 & -4.93 $\pm$ 0.14  \\ 
175640     & HR 7143    & B9 HgMn    &            &  12000 & 4.0   &  1.5 $\pm$    1 &  0.5 $\pm$  0.5 &              & -5.65 $\pm$ 0.12 & -4.98 $\pm$ 0.11  \\ 
175687     & $\xi^1$ Sgr & A0II       &            &   9400 & 2.3   &    9 $\pm$    1 &  2.5 $\pm$  0.3 &              & -6.60 $\pm$ 0.14 & -4.80 $\pm$ 0.14  \\ 
178065     & HR 7245    & B9 HgMn    & SB1        &  12300 & 3.6   &  2.0 $\pm$    1 &  0.5 $\pm$  0.5 &              & -6.02 $\pm$ 0.11 & -4.65 $\pm$ 0.14  \\ 
178329     & HR 7258    & B3V        & SB1        &  16600 & 4.2   &    7 $\pm$    1 &   nd $\pm$   nd & -4.27 $\pm$ 0.14 & -6.63 $\pm$ 0.18 & -4.94 $\pm$ 0.16  \\ 
181470     & HR 7338    & A0III      & SB2        &  10085 & 3.92  &  2.5 $\pm$  0.5 &  0.5 $\pm$  0.5 &              & -6.62 $\pm$ 0.14 & -4.82 $\pm$ 0.12  \\ 
182835     & $\nu$ Aql  & F2Iab      &            &   6800 & 2.0   &   10 $\pm$    1 &  5.0 $\pm$  1.0 &              & -6.50 $\pm$ 0.14 & -4.80 $\pm$ 0.14  \\ 
185395     & $\theta$ Cyg & F4V        &            &   6725 & 4.45  &    7 $\pm$    1 &  2.0 $\pm$  0.3 &              & -6.55 $\pm$ 0.12 & -4.80 $\pm$ 0.12  \\ 
186122     & 46 Aql     & B9 HgMn    &            &  12900 & 3.7   &  1.0 $\pm$  0.5 &  0.5 $\pm$  0.5 & -4.60 $\pm$ 0.14 & -7.65 $\pm$ 0.18 & -4.05 $\pm$ 0.14  \\ 
189849     & 15 Vul     & A4III m    & SB1        &   7850 & 3.7   &   13 $\pm$  1.5 &    4 $\pm$    1 &              & -6.60 $\pm$ 0.22 & -4.60 $\pm$ 0.22  \\ 
193452     & HR 7775    & A0 HgMn    & SB1        &  10600 & 4.15  &  1.0 $\pm$  1.0 &  0.5 $\pm$  0.5 & -4.30 $\pm$ 0.11 & -5.65 $\pm$ 0.12 & -4.17 $\pm$ 0.12  \\ 
209459     & 21 Peg     & B9.5V      &            &  10300 & 3.55  &  3.8 $\pm$  0.5 &  0.4 $\pm$  0.4 & -4.55 $\pm$ 0.11 & -6.42 $\pm$ 0.11 & -4.70 $\pm$ 0.11  \\ 
209625     & 32 Aqr     & A5 m       & SB1        &   7700 & 3.75  &  4.5 $\pm$  1.0 &  4.0 $\pm$  1.0 & -5.00 $\pm$ 0.22 & -6.30 $\pm$ 0.14 & -4.38 $\pm$ 0.14  \\ 
214994     & o Peg      & A1V        &            &   9500 & 3.62  &    7 $\pm$    1 &  2.0 $\pm$  0.3 & -4.75 $\pm$ 0.14 & -6.15 $\pm$ 0.14 & -4.35 $\pm$ 0.12  \\ 
\hline 
\end{tabular} 
\end{center} 

\end{table*}

The stars from this observing programme that have been found to have
\vsi\ values below about 15~\kms, so that velocity field effects are
potentially detectable in the line profiles, are listed in
Table~\ref{all-good-stars}. This table provides two names for each
star, a spectral type (generally obtained through the Simbad
database), an indication of binarity (either from the literature or
from our own observations), \te\ and \logg\ (obtained by using
available Str\"{o}mgren $uvby$ and/or Geneva photometry together with
the FORTRAN calibration programmes described by Napiwotzki,
Sch\"onberner \& Wenske (\cite{Napietal93}: $uvby$) and by K\"unzli,
North, Kurucz \& Nicolet (\cite{Kuenetal97}: Geneva)). The final
columns contain results from our own analysis that will be described
below.

Most stars in the survey were of course found to have values of \vsi\
large enough that we are not able to detect directly the local
velocity fields in the rotationally broadened line profiles. For such
stars, we nevertheless determined a few abundances, and accurate \vsi\
and sometimes $\xi$ values.  Some results for these stars are reported
in Table~\ref{all-good-stars}, but these stars are not discussed
individually any further in this paper.

\section{Treatment of double-line spectroscopic binaries}

Several of the stars in the present sample, noted as SB2 in
Table~\ref{all-good-stars}, are observed to be double-lined
spectroscopic binaries. To study the primaries of such systems, it
is very useful to subtract the secondary spectrum from the observed
composite spectrum. This is not essential for determination of the
rotation velocity of the primary (as long as we are careful to avoid
lines significantly blended by the secondary star), but determination
of the microturbulence parameter $\xi$ rests on line-by-line abundance
measurements, which in turn require that the line depths of the
primary spectrum be correctly normalised.

In the cases of greatest interest, the secondary is found from neutral/ion line
ratios to be a mid or late A spectral type of very low \vsi. We have
previously found (Landstreet \cite{jdl98}) that the line profiles of
the stronger lines in such stars are not correctly described by the
standard line model with thermal broadening, damping, and
height-independent Gaussian microturbulence. Thus, it is not practical
to compute a synthetic spectrum of the secondary which we could
subtract from the primary.  Furthermore, for none of the systems
studied here do we have enough information available (either from the
literature or from our own data) to be able to determine accurately
the secondary/primary light ratio in our short observational windows.

However, we have found that the secondary spectra in the four stars of
particular interest are all very similar to the observed spectra of
the very sharp-lined Am stars HD~108642 and HD~209625, previously
studied by Landstreet (\cite{jdl98}). In all cases the secondary
spectrum is almost unbroadened by rotation. This suggests that a
simple method of correcting approximately the composite spectra for
the effects of the secondary would be to Doppler shift an observed
spectrum of one of these two Am stars to coincide with the secondary
spectrum, subtract it, and renormalise the resulting spectrum.
Experiments have shown that this procedure in fact removes the
secondary spectrum from the observed composite spectrum, leaving a
residual noise in the difference spectrum which is roughly
proportional to the depth of the subtracted lines, and which is
generally no more than roughly 1--2\% of the continuum. This procedure
also allows us to identify with some certainty which lines of the
primary spectrum are significantly blended with secondary lines, so
that we can avoid including these lines in our detailed analysis of
profile shapes.

In general, it would be necessary to Doppler shift each pixel of the
``template'' secondary star individually, and then resample the
template star to the pixel spacing of the composite spectrum. In our
particular case, however, we have spectra of either HD~108642 or
HD~209625 taken with exactly the same instrumental settings used for
the SB2 which is to be corrected, and it is found by experiment that
it is adequate to simply shift the entire template spectrum by the
integer number of pixels required for the mean Doppler shift, and
subtract the template from the composite spectrum pixel by pixel. This
is possible because (a) the spectrum windows used are
sufficiently short (less than 100~\AA\ at 4500~\AA) and (b) because
the pixel spacing of the data is small enough (about 0.016~\AA). Point
(a) ensures that the error in the Doppler shift at the ends of the
window is small, and point (b) allows a pixel-to-pixel correspondence
to be found for which the fit of the template to the composite is very
good. 

\begin{figure*}
\begin{center}
\resizebox{16.0cm}{!}{
\includegraphics*[angle=0]{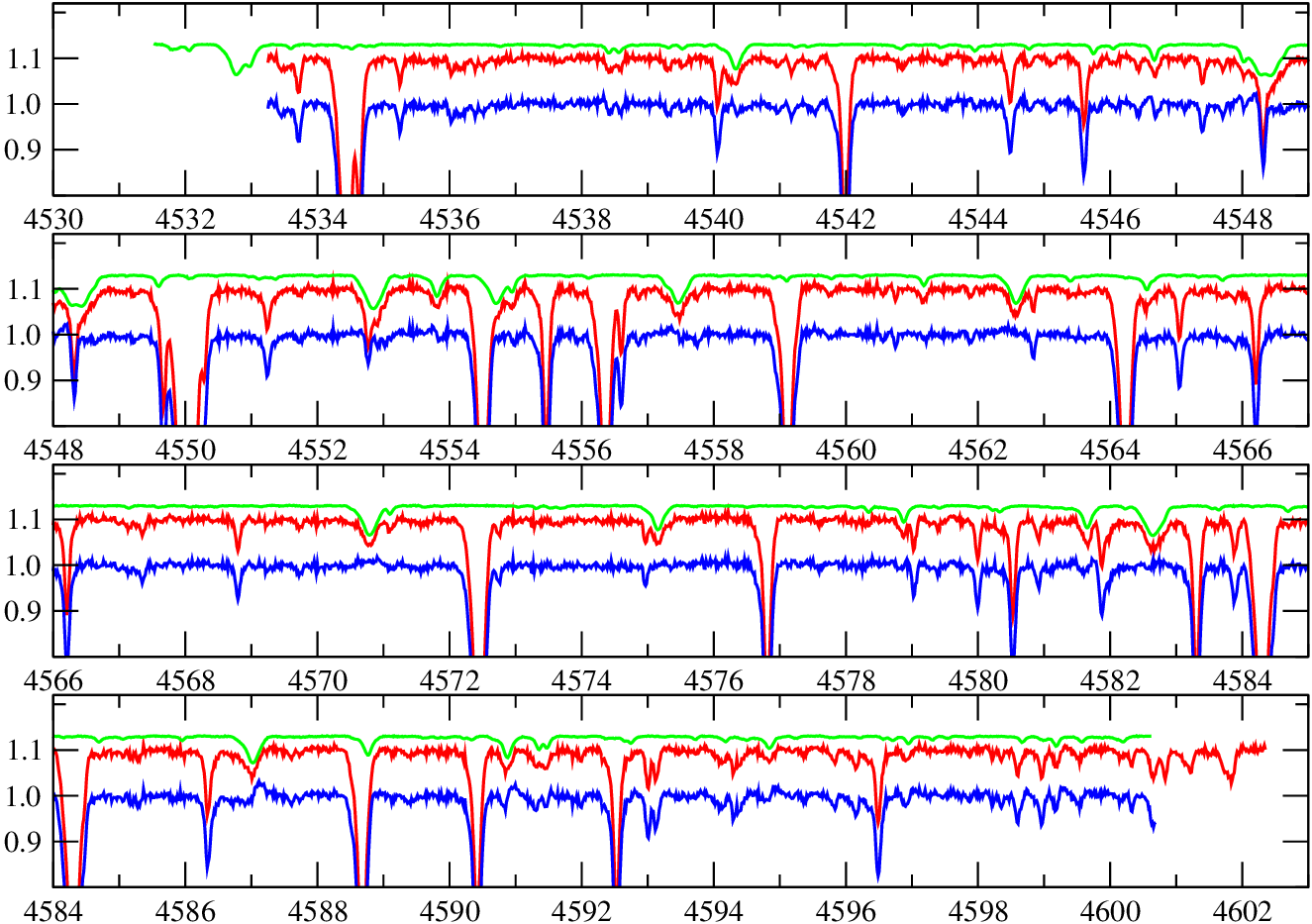}}
\caption{\label{hd103-hd108.fig} An example of subtraction of the
  secondary in an SB2 system. The figure shows the original spectrum
  of HD~103578 in the 4530~\AA\ window (middle spectrum, displaced
  upwards by 0.1 for clarity); the spectrum of HD~108642 renormalised
  and pixel-shifted to match the secondary in the HD~103578 system as
  well as possible (top spectrum, shifted upwards from initial
  spectrum by 0.03); and the final spectrum of HD~103578A produced by
  subtraction of the scaled and shifted spectrum of HD~108642 followed
  by renormalisation to a continuum value of 1.0 (lower spectrum).   }
\end{center}
\end{figure*}

We have therefore corrected each of the SB2 spectra analysed in detail
by shifting a template spectrum by an integer number of pixels,
rescaling the intensity of the template to fit the secondary spectrum
as well as possible, subtracting the template from the composite, and
rescaling the residual primary spectrum to return the continuum level
to 1.0. An example of this procedure is shown in
Figure~\ref{hd103-hd108.fig}.  It will be seen that the correction
  of the primary spectrum for individual spectral features of the
  secondary component is good but not perfect (especially near the red
  end of the spectrum); however, it is clear which lines in the
  primary are contaminated by lines of the secondary.

It is clear that this procedure does not ensure an accurate
determination of the light contribution of the secondary to the total
composite spectrum, but only an estimate of this contribution.
However, the uncertainty arising from this effect (essentially in the
determination of the zero point of the flux in the extracted primary
spectrum) is not very large. Typically it is found that the secondary
template star contributes about 10\% of the light to the composite.
Since the lines of the template stars, which always fit the profiles
of the secondary star rather accurately, are virtually unbroadened by
rotation and therefore quite deep (the Ba {\sc II} line at 4554.03
\AA\ descends to below 0.2 of the continuum), the amount of light
contributed by the secondary will not be overestimated by more than
about 10\% at the worst, or about 1\% of the total flux in the
composite spectrum. Similarly, if we assume that the secondary in the
composite SB2 spectrum has at least roughly solar composition, its
strongest lines, which are highly saturated, should be about as deep
as those of the template spectrum. Again, it appears that a 10\%
underestimate of the secondary's contribution to the total light is
unlikely, which again leads to an overall uncertainty after
subtraction of roughly 1\% of the continuum in the residual spectrum.
This is comparable to the uncertainty caused by correction
for scattered light in the composite spectrum, and will not lead to a
substantially increased error in either abundance determinations or in
the microturbulence parameter $\xi$.

The case of HD~114330 requires individual discussion. From speckle observations
there is a visual secondary about 0.45~arcsec from the primary,
approximately 2~mag fainter than the primary, and of similar or
slightly higher \te\ (e.g. Ten Brummelaar et
al. \cite{TenBetal00}). However, there is no trace of this component
in the observed spectrum. Because image slicers were used in both
spectrographs with which data were obtained for this project, light
from this component almost certainly is part of the observed
spectrum. Thus we suspect that the secondary spectrum is absent either
because its lines are extremely broad, or because they are extremely
narrow and coincide with those of the primary. We see no trace of
broad lines in our (blue) spectra, so most probably the lines of the
secondary are very narrow, and we must be concerned about how they
influence our analysis. If weak secondary lines are present on the
blue wing of each primary line, these contaminating lines could
perhaps produce excess blue wing absorption in strong lines (or in
most lines), and line profiles that do not closely match computed
ones. It is not clear that such contamination could lead to strong
lines that require a larger value of \vsi\ for best fit compared to
weak lines. Clearly, caution must be exercised in interpreting
spectra of HD~114330, and our conclusions cannot rest on this star
alone. 

\section{Modelling of spectra with ZEEMAN}

To uncover the information encoded in the observed spectral lines of a
star about the atmospheric velocity field(s) present, we must compare
the observed spectrum to a computed spectrum which incorporates some
model velocity fields, typically thermal broadening, the rotational
velocity \vsi, and the standard model of height-independent Gaussian
microturbulence. The first step in this process, so that we can select
an (approximately) appropriate atmosphere model from a precomputed
grid of ATLAS 9 solar composition models (Kurucz \cite{Kuru79}, \cite{Kuru92}), is to
determine the effective temperature \te\ and gravity \logg.

These parameters were determined using available
Str\"{o}mgren $uvby{\rm H}\beta$ and Geneva six-colour photometry,
together with appropriate calibrations. Temperatures determined using
the two different photometry systems usually agree within roughly 1\%
(typically 100~K at $\te \sim 10^4$~K), although of course the absolute
values of \te\ are probably only known to within 2--3\%. The agreement
of the inferred values of \logg\ was worse, typically 0.2--0.3 dex,
presumably because the Geneva system does not have a narrow-band filter
equivalent to the H$\beta$ filter of the Str\"{o}mgren system, and thus
is less sensitive to gravity (for hot stars) and more sensitive to
interstellar reddening (for cool stars). We generally adopted a value
of \logg\ close to the value given by the Str\"{o}mgren calibration.

For SB2 stars, approximate values of \te\ and \logg\ were found using
the combined light. We tried to improve these values by varying the
assumed values of \te\ and \logg\ to see if the quality of the fit to
our spectral windows was significantly improved by a somewhat
different choice of these fundamental parameters, but for all the
stars of interest we were unable to determine more accurately the
fundamental parameters of the primary star from our short spectra. We
therefore retained the estimates obtained using the combined light,
which in general probably yield values of \te\ that are slightly too
high, perhaps by 1--2\%.

The observed spectra of the stars in our sample have been modelled
using the FORTRAN programme zeeman.f (version zabn4.f), described by
Landstreet (\cite{jdl88}) and Landstreet et al.  (\cite{Landetal89}),
and compared to other magnetic line synthesis codes by Wade et
al. (\cite{Wadeetal01}). This is nearly the same programme that was
used in the previous study of line profiles of sharp-line A and B
stars by Landstreet (\cite{jdl98}) except for extensive modifications
to include the effects of anomalous dispersion for a stellar
atmosphere permeated by a magnetic field. These modifications have
essentially no impact on the analysis of the present (non-magnetic)
stars.

Zeeman.f makes extensive use of spectral line data obtained from the
Vienna Atomic Line Database (Kupka et al. \cite{Kupetal00}, \cite{Kupetal99};
Ryabchikova et al. \cite{Ryaetal97}; Piskunov et al. \cite{Pisetal95}).

A key feature of this programme is that it can be instructed to search
for an optimum fit to a single chemical element in one (or several)
observed spectra; at the same time, it finds the best fit radial
velocity for the spectrum, and the best fit value of \vsi. This
fitting is done assuming the standard model of height-independent
Gaussian microturbulence, and the fitting sequence can be carried out
for a series of different values of microturbulence parameter $\xi$ to
enable the best fitting $\xi$ to be selected. The programme returns
the value of the $\chi^2/\nu$ of the best fit ($\nu$ is the number of
degrees of freedom of the fit), as well as the parameters of the best
fit.

Zeeman.f can be used in several somewhat different modes. Two modes
were mostly used in this study. In one, we optimise the fit to an
element using all the significant lines of that element in one entire
observed spectrum (30 to 75~\AA\ long), for a series of values of
$\xi$, selecting in the end the value of $\xi$ which fits the spectrum
best, and the corresponding abundance. This method only leads to a
determination of $\xi$ if the spectrum includes spectral lines having
a range of strengths, at least some of which are saturated. If all the
spectral lines are relatively weak, different values of $\xi$ fit the
spectrum roughly equally well, up to the value at which
microturbulence broadens the line by more than its total observed
width. 

In the second mode, we optimise the fit of a set of selected spectral
lines of one element, one line at a time (i.e. in short windows,
typically 1--2~\AA\ wide, around each of the selected lines), as a
function of $\xi$, and then plot "Blackwell diagrams" (Landstreet
\cite{jdl98}): inferred abundance as a function of $\xi$ for each
spectral line separately. If some of the spectral lines used are
fairly saturated while others are weak, the resulting curves will
usually intersect in a well-defined small region, which determines
both the best fitting abundance of the element, and the best fit value
of $\xi$. The programme output using this method allows us to examine
conveniently the quality of the fit of an individual model spectral
line to a particular observed line as a function of the various
parameters such as \vsi\ and $\xi$, and makes it easy to see whether
or not the same fitting parameters are derived from all spectral lines
of one (or several) element(s). Some Blackwell diagrams are shown by
Landstreet (\cite{jdl98}).

The available high-quality spectra for each star have been analysed as
described above. In general, we have optimised abundances of the
principal detectable elements (some of He, O, Si, Ti, Cr, Fe, Ni, Zr,
Ba) as a function of the microturbulence parameter $\xi$, and chosen
the microturbulence parameter value which yielded the best concordance
of abundances derived from different spectral lines of one to three
elements (usually Ti, Cr, and Fe, and sometimes Si) for which enough
lines, with a large enough range of equivalent width, are
available. The deduced value of $\xi$ is then used to determine
abundances of all other elements.

The best-fit values of \vsi\ and $\xi$, and the abundances of Si, Cr,
and Fe, with our estimates of uncertainties, are reported in Table
\ref{all-good-stars}. In cases where the available range of line
strengths in our spectra was not sufficient to allow us to determine
the value of $\xi$ we report ``nd $\pm$ nd''
(for ``not detected''). In a number of stars we were
only able to obtain upper limits to the values of $\xi$, usually from
Blackwell diagrams, but occasionally from the very narrow line
profiles; such cases are reported with equal values and uncertainties
for $\xi$ (e. g. $0.5 \pm 0.5$ \kms). 

Abundances are given as logarithmic number densities relative to H,
because this is essentially what we measure; for many stars in our
sample the atmospheric abundance of He is unknown (and probably not
close to solar), so we prefer to report $\log(n_X/n_{\rm H})$ rather
than $\log(n_X/n_{\rm tot})$ values, which requires an {\it ad hoc}
assumption about $n_{\rm He}/n_{\rm H}$. Uncertainties in the quoted
abundance values are due in part to fitting uncertainties, which arise
because abundances derived from individual spectral lines for a given
value of $\xi$ always show some scatter about the average, due to
imprecise atomic data (especially $gf$ values), to the fact that we do
not have a correct model of strong lines with which to fit the
observations, and to the fact that the atmosphere model used is only
an approximate description of the actual stellar atmosphere. These
fitting uncertainties are typically of the order of 0.1~dex in
size. There is an additional uncertainty from the fact that the
fundamental parameters of each star are only known with a limited
precision. We assume that the value of \te\ has an accuracy of
approximately $\pm 3$\%, and that \logg\ is known to about $\pm
0.25$~dex. We have carried out a number of experiments to test the
effect of these uncertainties on the derived abundances, by rederiving
abundance values after changing \te\ or \logg. Because our abundances
are generally obtained from the dominant ion, the abundances are not
very sensitive to changes in either \te\ or \logg. The uncertainties
in fundamental parameters change the derived abundances by an amount
which varies from element to element but is of the order of
0.1~dex. The total uncertainty given for each abundance in Table
\ref{all-good-stars} is then the sum in quadrature of the individual
fitting uncertainty and a fixed uncertainty of 0.1~dex arising from
fundamental parameter uncertainties.

Note also that since we do not model macroturbulence, but include all
broadening except for instrumental, thermal, and microturbulent in our
\vsi\ broadening, the values of \vsi\ for the coolest stars of our
sample are larger than those given in studies which separate the
rotational and macroturbulent components of line broadening.

For the stars with the sharpest lines, the precise value of $\xi$
obtained by our fitting procedure may not be the same as the value
found by other methods, even with the same observational material and
atomic data. This is because we fit line {\em profiles} using a
$\chi^2$ minimisation procedure rather than fitting equivalent
width. As will be seen below (and was discovered by Landstreet
\cite{jdl98}), for the sharp-line A stars the standard line profile
model with microturbulence does not fit the profiles of some very
strong lines at all well, effectively showing that this is an
inadequate description of these lines. In this situation, it is not
clear that there is any good reason to prefer one fitting method over
the other -- but it may well be that in some cases they yield slightly
different values of $\xi$.

We have compared the basic parameters we have chosen for stars
analysed, and the resulting abundance values, with a number of
previous analyses available in the literature (e.g. Hill \& Landstreet
\cite{HillLand93}; Adelman et al. \cite{Adeletal97}; Hui-Bon-Hoa
\cite{Hui00}; Pintado \& Adelman \cite{PintAdel03}; Gebran et
al. \cite{Gebretal08}). In most cases our value of \te\ is within
about 200~K, and our \logg\ within about 0.2~dex, of values adopted by
other authors. Our values of $\xi$ are mostly in agreement to within
0.5~\kms\ or less, and our resulting element abundances agree within $\pm
0.2$~dex. These discrepancies are fairly typical for the differences
between independent abundance analyses of B and A stars, and probably
derive from differences in fundamental parameters, detailed model
atmospheres used, choices of $gf$ values and wavelength regions,
etc. The only significant systematic disagreement we find is that the
element abundances we derive for the supergiants in our sample are
almost always of order -0.3~dex lower than the LTE abundances found by
Venn (\cite{Venn95}). The origin of this difference has not been
identified, but does not affect the analysis of spectral line profiles
which is the main subject of this paper.

\section{Evidence of atmospheric velocity fields}

\begin{figure}
\begin{center}
\resizebox{8.0cm}{!}{
\includegraphics*[angle=0]{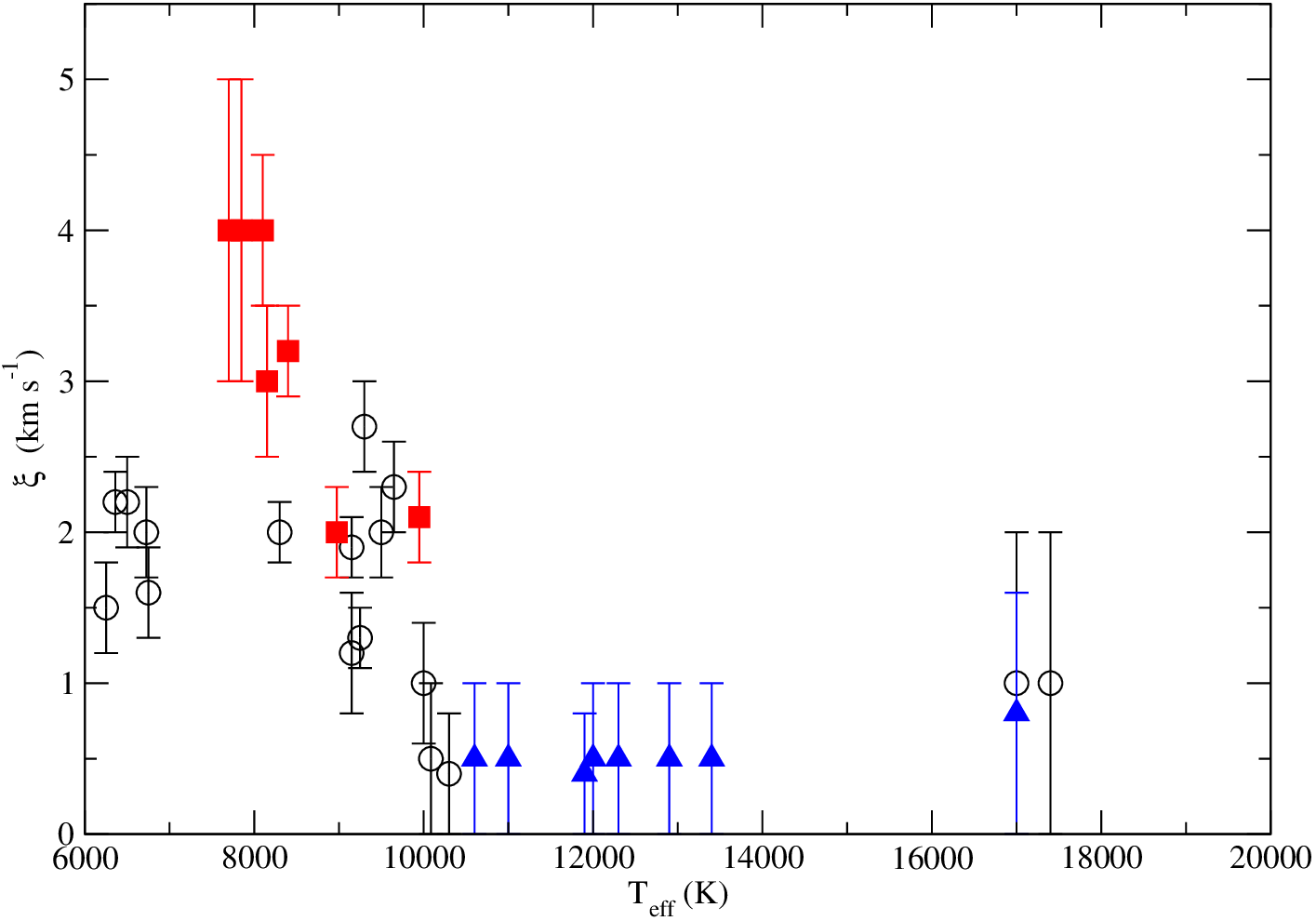}}
\caption{\label{xi-vs-teff.fig} Microturbulence parameter $\xi$ as a
  function of stellar \te\ for main sequence stars in the sample. Open
  symbols are normal stars, filled squares are Am stars, and filled
  upright triangles are HgMn stars.  }
\end{center}
\end{figure}

In our previous study of atmospheric velocity fields in middle main
sequence stars (Landstreet \cite{jdl98}), several kinds of
observational evidence of velocity fields were studied: strong lines
whose profiles departed significantly from those predicted by the
standard microturbulence model; significant (in some cases quite
strong) excess absorption in the blue wings of strong lines; deduced
rotation velocities \vsi\ which increase with line equivalent width;
and the microturbulence parameter $\xi$. Microturbulence has been
known and studied for decades, but the other spectral line symptoms of
local velocity fields were systematically studied in tepid main
sequence stars for the first time in our previous study.

One of the results of the spectrum synthesis carried out for the stars
of this sample is a determination of the microturbulence parameter
$\xi$ (see Table \ref{all-good-stars}) for all the stars of the sample
except a few hot stars that have no saturated lines in the window(s)
observed. The derived values of $\xi$ are plotted as a function of
stellar effective temperature \te\ in Figure \ref{xi-vs-teff.fig},
with separate symbols for normal, Am, and HgMn stars. This figure
clearly shows the expected behaviour: $\xi$ is zero within the
uncertainties for $\te \geq 10\,000$~K, while below this temperature
it rises rapidly towards lower \te\ to a limiting value which for most
stars is approximately 2~\kms. 

The peculiar stars of this diagram have quite distinctive behaviour. In
none of the Ap HgMn stars is a non-zero value of $\xi$ detected,
consistent with models of these stars that require vertically
stratified atmospheres (e.g. Sigut \cite{Sigu01}). In contrast, all
the metallic-line Am stars have clearly non-zero values of $\xi$. For
the hotter Am stars ($\te \ge 8\,500$~K), the value of $\xi$ is
indistinguishable from that of normal stars of the same \te\ value, but
the cooler Am stars tend to have $\xi$ values about 1 to
2~\kms\ larger than the values found for normal stars. The relatively
large $\xi$ values found for the five cool Am stars of the present
sample are consistent with the results of Landstreet (\cite{jdl98})
for a smaller sample.

\begin{figure}
\begin{center}
\resizebox{9.0cm}{!}{
\includegraphics*[angle=0]{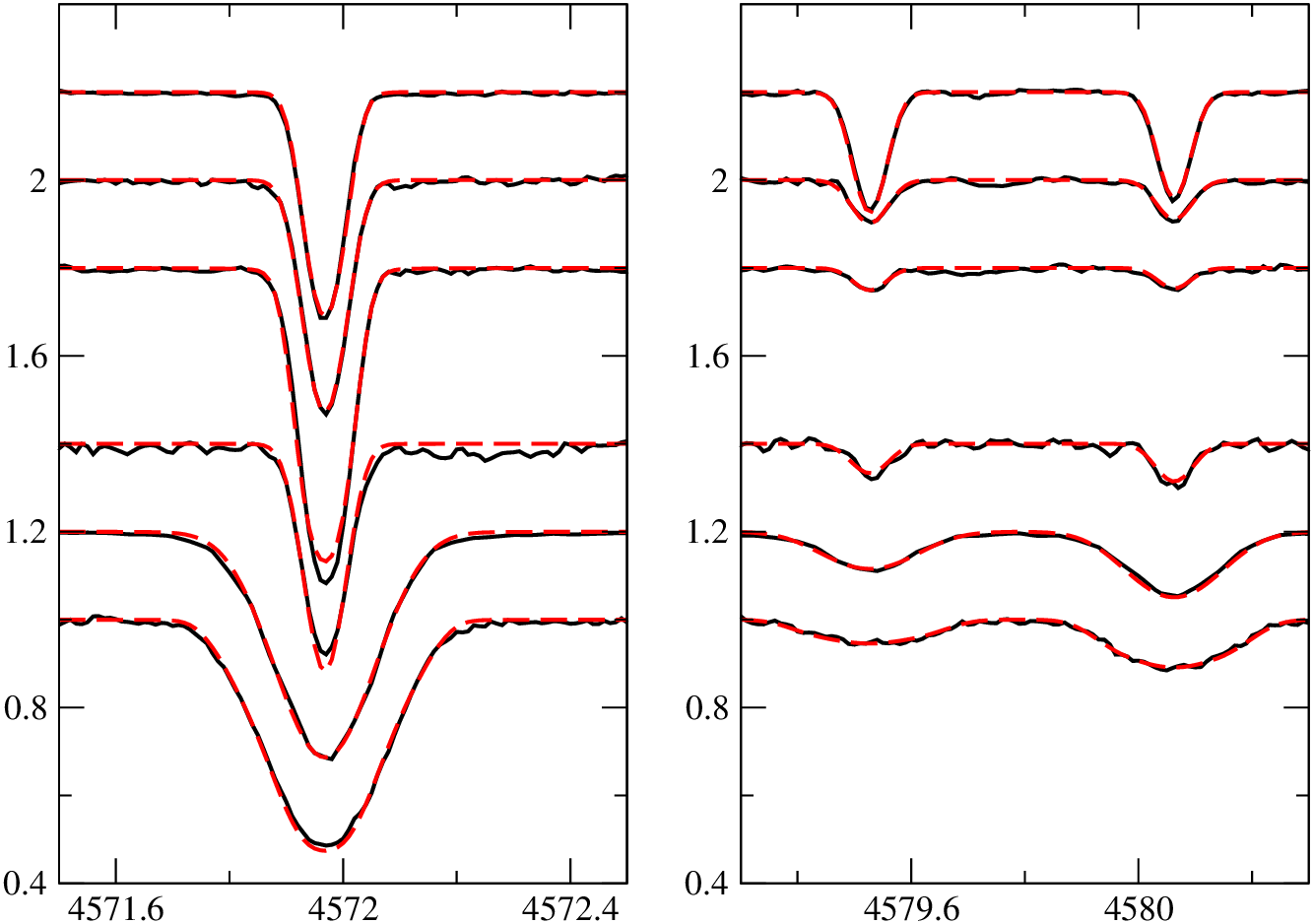}}
\caption{\label{ti4572-fe4580.fig} At most very weak convection
  signatures in spectral lines of stars with \te\ values between 13000
  and 9000~K observed in the 4530 -- 4600~\AA\ window. From top to
  bottom, in order of decreasing \te: HD~186122 (46 Aql), HD~178065,
  HD~175640, HD~181470A, HD~214994 (o Peg, exceptionally observed only
  with $R = 40\,000$), and HD 73666. The top three stars are late B
  HgMn stars, while the lower three stars are normal A stars (one of
  which is an SB2). The solid lines show the observed line profiles of
  the strong 4571.97~\AA\ line of Ti {\sc ii} (left panel) and two
  weak to moderate lines of Fe {\sc ii} at 4579.53 and
  4580.06~\AA\ (right panel); the dashed lines show profiles computed
  with the standard line model of depth-independent, isotropic
  microturbulence, using the parameters of
  Table~\ref{all-good-stars}. }
\end{center}
\end{figure}

The most striking symptom of atmospheric velocities discovered by
Landstreet (\cite{jdl98}) was the large departure of the line profiles
of strong spectral lines in a few stars from the profiles predicted by
the standard model of a height-independent, isotropic Gaussian
(microturbulent) velocity field. The velocity field signature takes
the form of line profiles of strongly saturated lines that are much
more triangular than line profiles predicted by the standard model.
Furthermore, these strong lines have substantial excess absorption in
the blue line wing relative to synthetic line profiles. Such pointed,
asymmetric profiles are reminiscent of the line profiles found in
slowly rotating cool stars, which are modelled by introducing
macroturbulence. It is clear that such line distortions signal the
presence of local atmospheric motions more complex than those
assumed by the standard microturbulent model, and provide valuable
information about these motions. 


\begin{table*}[ht]
\begin{center}
\caption{\label{detect-vels} Main sequence A-type stars in which evidence of velocity fields is directly detected in line profiles}
\begin{tabular}{llllllllll}
\hline\hline
  HD     &  other    & spectral & binarity & \te & $\xi$  & depressed & line shape  & \multicolumn{2}{c}{\vsi} \\ \cline{9-10}
         &           & type     &          &     &        & blue wing & discrepancy & weak lines & strong lines \\
         &           &          &          & (K) & (\kms) &           &             & (\kms)    &  (\kms)     \\
\hline
2421 A   & HR 104 A  & A2Vs &       SB2 &  10000 & 1.0    &    weak   & none        & 3.5       & 3.5 \\
72660    & HR 3383   & A1V  &           &  9650  & 2.3    &    weak   & weak        & 5.0       & 6.1 \\
114330   & $\theta$ Vir & A1Vs &    VB2 &  9250  & 1.3    &    weak   & weak        & 4.0       & 5.5 \\
157486 A & HR 6470 A & A0V  &       SB2 &  9150  & 1.9    &    weak   & weak        & 1.5       & 3.5 \\
27962    & 68 Tau    & A2IV m &     SB1 &  8975  & 2.0    &    weak   & weak        & 11        & 11.5 \\
103578 A & 95 Leo A  & A3V &        SB2 &  8300  & 2.0    &    strong & strong      & 3         & 9 \\
107168   & 8 Com     & A8 m &           &  8150  & 3.0    &    weak   & weak        & 13        & 14 \\
108642 A & HR 4750 A & A2 m &       SB1 &  8100  & 4.0    &    strong & strong      & 2         & 10 \\
209625   & 32 Aqr    & A5 m &       SB1 &  7700  & 4.0    &    strong & strong      & 4.5       & 10 \\
\hline\hline
\end{tabular}
\end{center}

\end{table*}


In the previous study, the phenomenon was only observed in very clear
form in two stars (HD~108642 and HD~209625) with \te\ values close to
8000~K, both of which are Am stars. One further normal early A star
(HD~72660) was found to show a very attenuated departure of observed line
profiles from computed ones, primarily in the form of observed
blue line wings that are slightly depressed below the computed line
wings. For the other two stars in that study with \te\ below
10\,000~K, the \vsi\ values are too large for a significant departure
from a purely rotational profile to be seen. All observed line
profiles of three very sharp-line stars hotter than $\te \approx
10\,000$~K were found to be in very good agreement with the computed
profiles; no indication of any atmospheric velocity field other than
bulk stellar rotation is seen in the lines.  As discussed in the
introduction, the main goals of the present project are to find more
examples of such line profile discrepancies, to discover whether the
phenomenon is restricted essentially to Am stars, or is found more
generally, and to determine more clearly over what range of
\te\ directly observable velocity field symptoms are found. 

\begin{figure}
\begin{center}
\resizebox{9.0cm}{!}{
\includegraphics*[angle=0]{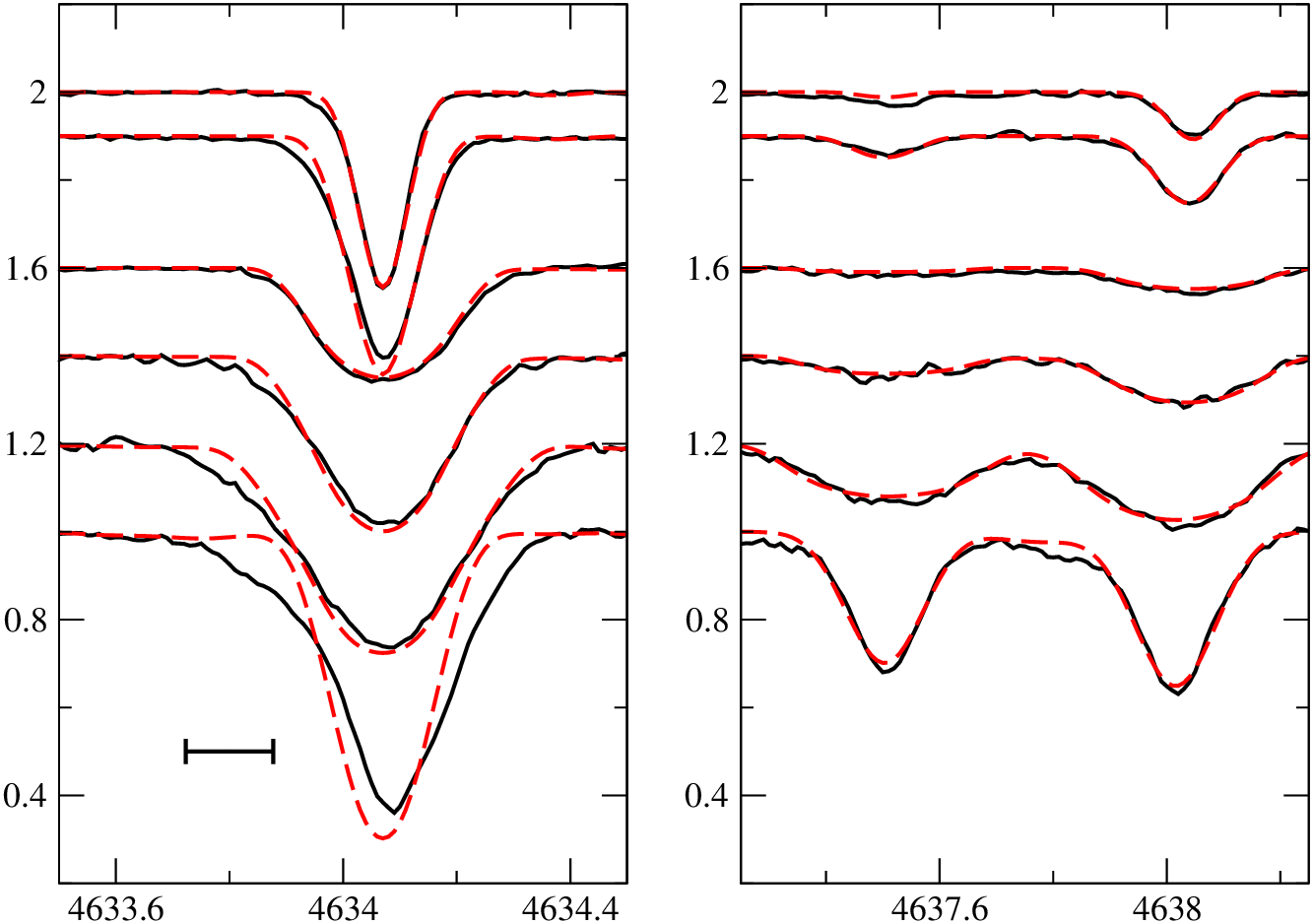}}
\caption{\label{cr4634-fe4638.fig} Convection signatures (particularly
  the depressed blue line wing) in spectral lines of stars with
  \te\ in the range 10000 to 7500~K, observed in the 4615 --
  4645~\AA\ window. From top to bottom, in order of decreasing \te:
  HD~2421A, HD~72660, HD~47105, HD~27962, HD~107168, and
  HD~209625. The top three stars are normal A stars, while the lower
  three stars are all Am stars. The solid lines show the observed line
  profiles of the strong 4634.07~\AA\ line of Cr {\sc ii} (left panel)
  and the weak 4637.50~\AA\ line of Fe {\sc i} and the blend of
    4638.01~\AA\ of Fe {\sc i} and 4638.05~\AA\ of Fe {\sc ii}
  (right panel); the dashed lines show profiles computed with the
  standard line model of depth-independent, isotropic microturbulence,
  using the parameters of Table~\ref{all-good-stars}. The bar in
    the lower left corner of the left panel represents a wavelength
    shift equivalent to 10~\kms.}
\end{center}
\end{figure}

\begin{figure}
\begin{center}
\resizebox{9.0cm}{!}{
\includegraphics*[angle=0]{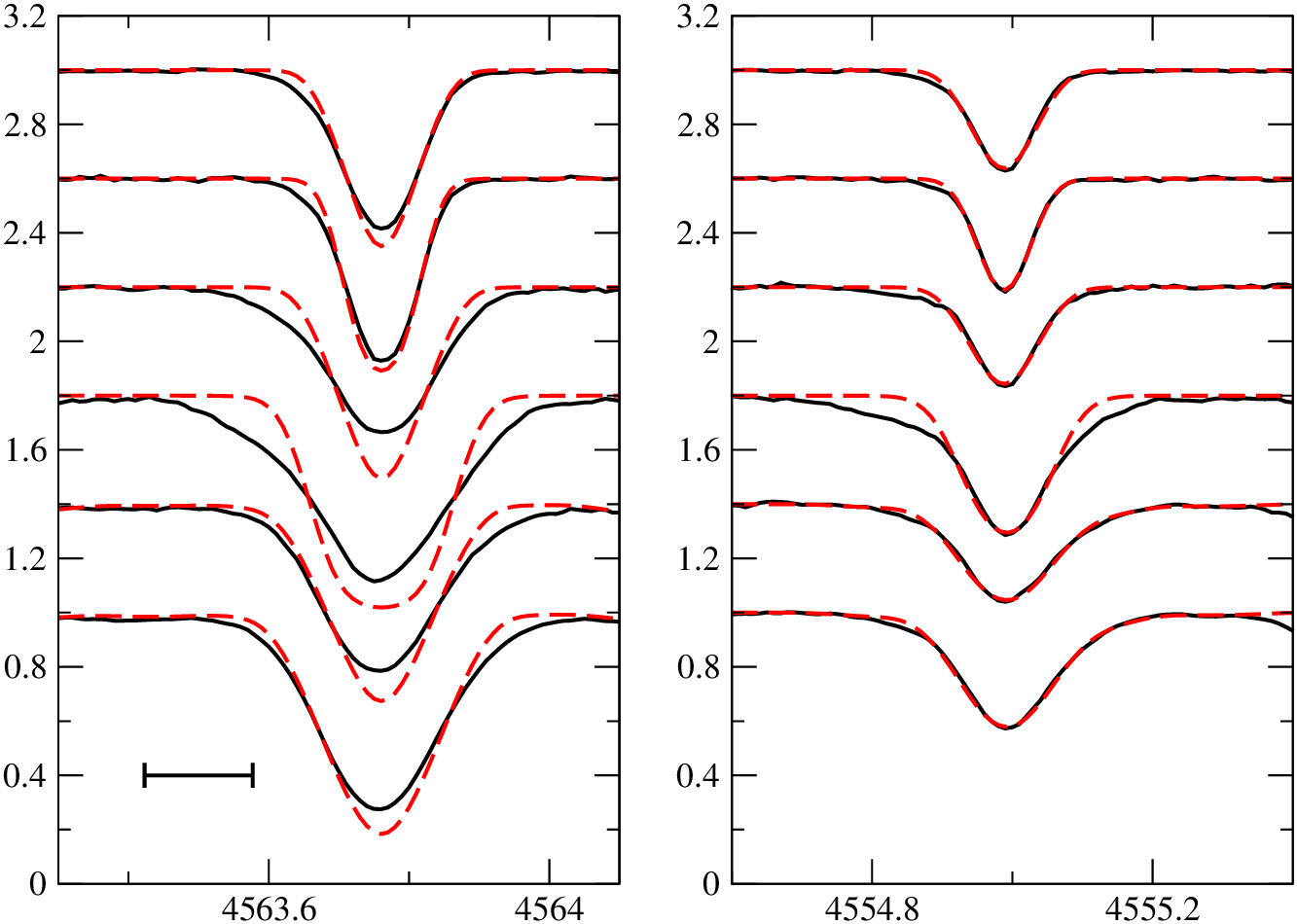}}
\caption{\label{ti4564-cr4555.fig} Convection signatures in spectral
  lines of stars with \te\ in the range 9250 to 6500~K, observed in
  the 4530 -- 4600~\AA\ window. From top to bottom: HD~114330,
  HD~157486A, HD~103578A (95 Leo A), HD~108642A, HD~185395 ($\theta$
  Cyg), and HD~61421 (Procyon). The top star is a normal A star, the
  next two stars are normal (SB2) A stars, the fourth star is an Am
  star, and the bottom two are normal F stars. The solid lines show
  the observed line profiles of the strong 4563.76~\AA\ line of Ti
  {\sc ii} (left panel) and the weak to moderate 4554.99~\AA\ line of
  Cr {\sc ii} (right panel); the dashed lines show profiles computed
  with the standard line model of depth-independent, isotropic
  microturbulence, using the parameters of
  Table~\ref{all-good-stars}. The bar in the lower left corner of
    the left panel represents a wavelength shift equivalent to
    10~\kms.}
\end{center}
\end{figure}

In Figures \ref{ti4572-fe4580.fig} through \ref{ti4564-cr4555.fig} we
compare the result of detailed modelling of lines of the
sharpest-lined main sequence stars in the present sample with observed
line profiles. Two spectral windows are shown, reflecting the
available observational material for the stars of the sample: data in
the window around 4560~\AA\ were obtained in 2000--01, while data in
the window around 4630~\AA\ are from earlier observing runs. In each
figure, the line in the left panel is the strongest spectral line in
the observed window that is essentially free from blends in the range
of \te\ of the figure. The right panel shows a relatively weak line
that is also largely blend-free. These observed lines are compared
with model spectral lines computed assuming the standard model of
velocity broadening due to the atomic thermal velocities,
depth-independent, isotropic Gaussian microturbulence, and rigid-body
rotation, along with appropriate instrumental broadening.

As mentioned above, in stars showing significant line asymmetry, it is
also found that the \vsi\ value that best fits the strongest lines is
almost always significantly larger (by up to several \kms) than that
which best fits the weakest lines. Presumably this effect reflects the
extra contribution of the local macroscopic velocity field to the line
width of saturated lines, and the true value of \vsi\ is (at most) as
large as the minimum value found using the weakest, sharpest lines. In
the figures, the projected rotational velocities of the synthetic
lines with which these observed lines are compared are determined
using only the {\em weakest} lines, and the {\em same} \vsi\ value is used
for all lines of each star. The value of $\xi$ used is the value
reported in Table \ref{all-good-stars}, derived from the abundance
modelling of weak and strong lines.

These three figures clearly display the same phenomena identified by
Landstreet (\cite{jdl98}): no trace of macroscopic velocity fields
above about 10\,000~K, while below that temperature $\xi$ rises
rapidly from zero, and particularly sharp-lined stars show
\vsi\ values that depend on line strength, line asymmetries, and
significant discrepancies between computed and observed line profiles
of strong lines. 

For each of the four stars in Figure \ref{ti4572-fe4580.fig} with
effective temperatures above about 10\,000~K (HD~186122, 178065,
175640, and 181470A, the first three of which are Ap HgMn stars), the
value of $\xi$ is not significantly different from zero. The {\em
  same} value of \vsi\ fits weak and strong lines, and in all cases
the fit is very good to both weak and strong lines, which are clearly
symmetric. These stars show no significant trace of local large-scale
velocity fields such as ``macroturbulence''. The same result was found
for the normal star HD~209459 = 21 Peg ($\te = 10200$~K) by Landstreet
(\cite{jdl98}).

Of the seven stars in these figures with \te\ between 9\,000~K and
10\,000~K, the four with \vsi\ below 6~\kms\ (HD~2421A, HD~72660,
HD~114330, and HD~157486A) show small but clear deviations from the
synthesized line profiles, specifically depression of the observed
blue line wing relative to the computed line for the strong (but not
the weak) lines. In contrast, the two stars with $\vsi \sim
10$~\kms\ (HD~73666 and 47105), and the one star of $\vsi =
7$~\kms\ observed with $R = 40\,000$ show no detectable effect, even
though their $\xi$ values are comparable to those of HD~2421 and
HD~157486; for these stars the small velocities of the local velocity
field are concealed by the rotational and/or instrumental broadening.

There are six sharp-line stars in the sample with \te\ in the range
7\,500 to 9\,000~K: HD~27962, HD~103578A, HD~107168, HD~108642,
HD~189849, and HD~209625. (Two of these are the two Am stars
studied by Landstreet \cite{jdl98}.) In this temperature range the
local velocity field amplitude is large enough to be clearly visible
even with \vsi\ above 10~\kms\ (HD~27962 and 107168, but not in
HD~189849 for which our observational material is of low quality),
and, as previously found, in stars with sharper lines the impact of
the local velocity field is really striking. In HD~103578A, 108642 and
209625 the deviations from the model are clearly visible even in the
weak lines used to determine \vsi. In these stars the discrepancy
between the calculated and observed line profiles is very substantial
for the strong lines, both in the core shapes and in the occurrence of
observed blue line wings that are shallower and deeper than the
computed ones. In HD~27962 and 107168, both of which have \vsi\ values
above 10~\kms, the effects are visible but not striking.

Finally, we also show two F stars in Figure \ref{ti4564-cr4555.fig}
whose \te\ values are below 7\,000~K (HD~185395 and HD~61421), and
which show the well-known shape discrepancies of solar-type stars in
the strong lines, with the {\em red} line wings of the strong observed
lines shallower than and depressed below the model profile. Note that
the line profiles of A-type stars with depressed blue wings as shown
in Figure \ref{ti4564-cr4555.fig} are very different from the blueward
hook phenomenon which has been observed for F-type stars such as
Procyon and explained to originate from saturation of strong lines in
upflow regions (Allende Prieto et al. \cite{Prieetal02}). The cool
star blueward hook feature only affects lines close to their continuum
level and is much smaller in terms of velocities (less than 0.5~\kms).

To provide a scale for the velocity amplitude required to account
for the line profile distortions, in each figure where these effects
are apparent we include a small horizontal error bar with a total width of 
10~\kms.

A summary of all the main sequence A stars in which we find direct
evidence of an atmospheric velocity field (beyond that provided by a
non-zero value of $\xi$) has been found is given in Table
\ref{detect-vels}. Columns 3 to 6 recall the spectral type, binarity,
effective temperature and microturbulence parameter of these stars. In
the column ``depressed blue wing'' the comment ``weak'' or ``strong''
indicates the extent to which the short-wavelength line wing is deeper
than the long-wavelength wing. The column ``line shape discrepancy''
describes qualitatively the extent of discrepancies, other than the
depressed blue line wing, between observed and calculated line
profiles of the strongest spectral lines. The last two columns list
extreme values of \vsi\ as derived from best fits to the weakest and
strongest spectral lines in our spectra. As discussed qualitatively
above, we see that the difference between the best fit \vsi\ values
for the weakest and strongest lines in our spectra seems to increase
quite rapidly for \te\ below about 9000~K to a value that can be
several \kms\ at around 8000~K. (Note that HD~3883, noted by Gray \&
Nagel [\cite{gn89}] as having a reversed bisector, is omitted from
Table \ref{detect-vels} because we have no spectra of this star.)

We notice that the direct evidence of velocity fields in the line
profiles agrees very well with the indirect evidence provided by the
microturbulence parameter. Main sequence stars having \te\ above about
10000~K have $\xi$ values that are indistinguishable from 0, and thus
lack any evidence of small-scale velocity fields. These same stars
have spectral lines that are in excellent agreement with calculated
line profiles in which the broadening is assumed to be due only to
rotation, thermal broadening, collisional damping, and instrumental
resolution; these stars thus do not show any direct evidence of local
atmospheric hydrodynamic motions. 

In contrast, stars with \te\ below 10000~K all have non-zero values of
$\xi$, and all the stars with ``sufficiently'' small values of
\vsi\ also show direct line profile evidence of velocity fields. Local
velocity fields are detectable through differences between profiles of
observed strong spectral lines and computed lines, through asymmetries
in the observed lines, and in most cases through best-fit \vsi\ values
that are systematically larger for strong than for weak
lines. Furthermore, the strongest directly detected symptoms of local
velocity fields, observed in stars with \te\ around 8000~K, generally
coincide with the largest deduced values of $\xi$.

These results confirm the findings from a smaller dataset by Landstreet
(\cite{jdl98}), where it was shown that the range of effective
temperatures for which non-zero-values of $\xi$ are found coincides
rather closely with the occurrence of a convectively unstable region
in the stellar atmosphere as predicted by the Schwarzschild criterion
applied to the model atmosphere used for spectrum synthesis. Thus, the
occurrence of convective motions in the atmospheres is expected in
essentially the range of \te\ in which it is observed. The importance
of the observations presented in this paper comes from the direct
signatures of this convection in spectral lines, which should be able
to provide significant constraints on the nature of the actual flows
occurring in the atmosphere.

\begin{figure}
\begin{center}
\resizebox{9.0cm}{!}{
\includegraphics*[angle=0]{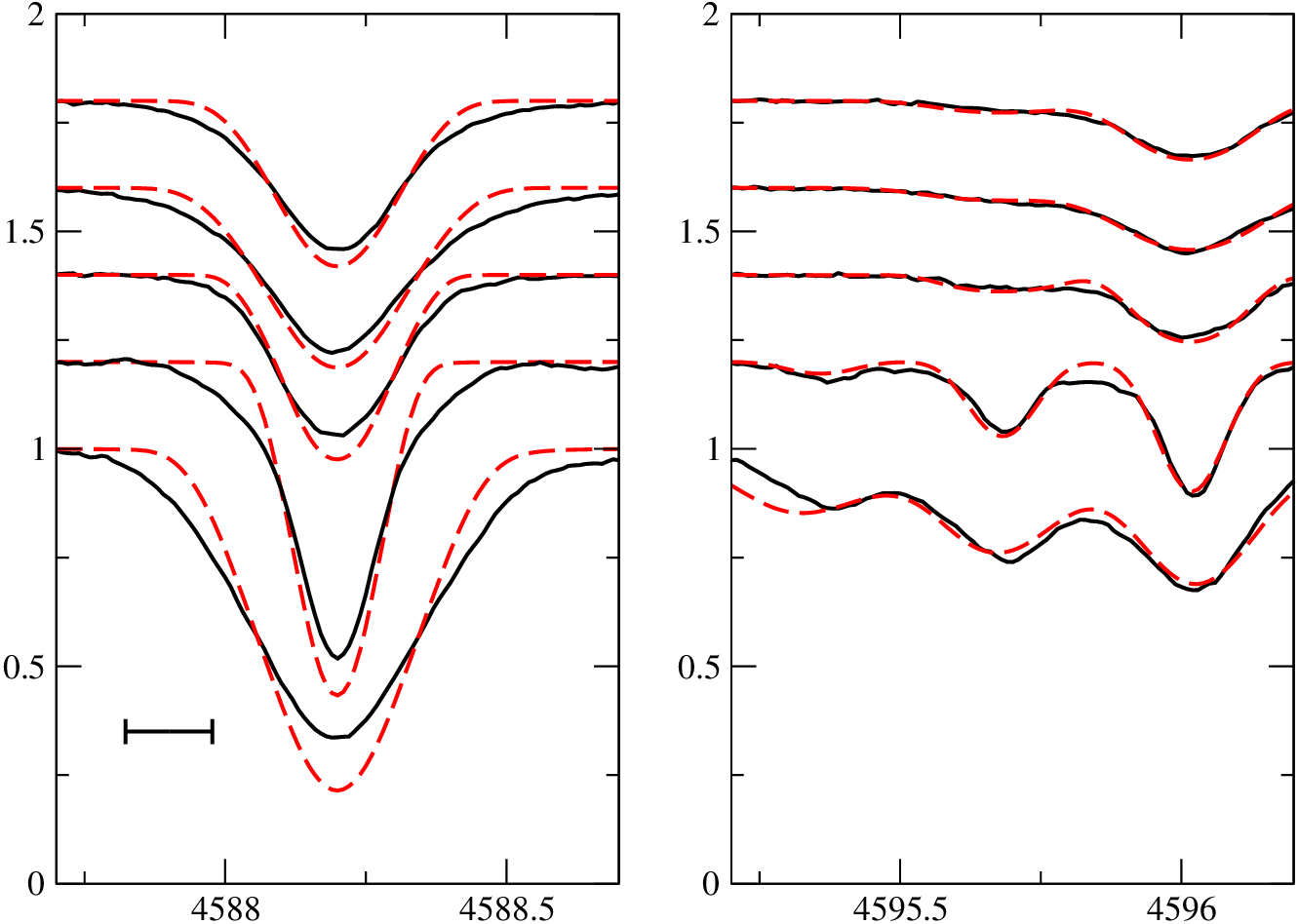}}
\caption{\label{cr4588-fe4596-sg.fig} Striking convection signatures
  in spectral lines of A and early F-type supergiants with
  \te\ values between 9700 and 6800~K observed in the 4530 --
  4600~\AA\ window. From top to bottom: HD~46300 (13 Mon), HD~87737
  ($\eta$ Leo), HD~175687 ($\xi^1$ Sgr), HD~147084 (o Sco), and
  HD~182835 ($\nu$ Aql). The solid lines show the observed line
  profiles of the strong 4588.20~\AA\ line of Cr {\sc ii} (left panel)
  and a group of weak to moderate lines, 4595.36 and
    4596.06~\AA\ of Fe {\sc i} and 4595.68 and 4596.02~\AA\ of Fe {\sc
      ii} (right panel); the dashed lines show theoretical profiles
  computed with the standard line model of depth-independent,
  isotropic microturbulence computed with the parameters of
  Table~\ref{all-good-stars}. The bar in the lower left corner of
    the left panel represents a wavelength shift equivalent to
    10~\kms.}
\end{center}
\end{figure}

It may be noted that the line profile distortions relative to the
canonical model are observed essentially only in lines of once-ionised
atoms such as Fe, Cr and Ti, and not in the few visible lines of
neutral atoms. This is presumably due to the fact that all neutral
lines are relatively weak, while the effects we are interested show
themselves in strong, heavily saturated lines.

It is of interest to compare the symptoms of local velocity fields
found in the spectra of main sequence stars with those found in lines
of supergiants of similar spectral types, and for this reason several
of the stars in our survey are such luminous objects. It is well known
that A and F supergiants require relatively large values of the
microturbulence parameter (typically several \kms) for successful
spectrum synthesis (e. g. Venn \cite{Venn95}; Gray et
al. \cite{Grayetal01}; Przybilla et al. \cite{Przyetal06}), and so we
expect to find strong evidence of velocity fields in the line
profiles of the slowest rotators. 

Five supergiants, ranging in \te\ from 9700 to 6800~K, are listed in
Table \ref{all-good-stars} and are shown in Figure
\ref{cr4588-fe4596-sg.fig}. As with the main sequence stars, the \vsi\
values have been determined using only the weakest lines, and for each
star the same value of \vsi\ has been used for all synthesized
profiles. A systematic result of our use of only the weakest lines is
that the \vsi\ values determined here are always significantly smaller
than those found in previous studies, where \vsi\ is fit mainly to the
stronger lines. In the right panel of the figure an example of
relatively weak lines shows that these profiles are fairly well fit by
simple rotational broadening (together with thermal and instrumental
broadening and classical microturbulence). In contrast, the observed
strong lines of the left panel are all significantly different in
shape from the computed lines. If we had (unphysically) allowed \vsi\
to be a free parameter for each spectral line separately, the strong
lines could have been substantially better fit. In each case the best
fit \vsi\ value determined from strong lines is a few \kms\ larger
than that found from weak lines, and is similar to values found in
other high-resolution studies. In the coolest star shown, HD~182835,
the best fit \vsi\ value found from weak lines is no more than
10~\kms, while the best fit for the strongest lines is about 18~\kms!

Thus, we find in these highly luminous stars evidence for the local
velocity fields phenomenologically similar to that found in main
sequence stars of similar \te, but with larger amplitude. However,
there is one significant difference: the supergiants do not show the
strong line asymmetry seen in the main sequence stars with similar
temperatures. The depression of the observed wings of strong lines
below the computed wings is essentially the same on the blue and red
sides of the line. This result is different from that of Gray \& Toner
(\cite{GrayTone86}), who find significant excess absorption in the
blue wings of spectral lines of F supergiants. However, their data
suggest that the asymmetry diminishes towards earlier spectral
types.  As their earliest supergiant has spectral type F5 while our
latest type is F2, perhaps we are simply seeing the asymmetry decrease
below the threshold for easy visibility.

A different possibility for the velocity field observed in supergiants
is that the line profiles reveal the presence of pulsations. Aerts et
al. (\cite{Aertetal08}) have shown that the qualitatively similar line
profiles of an early B-type supergiant, typically modelled by
macroturbulence, may be produced by numerous low-amplitude
gravity-mode oscillations, rather than by large-scale convective
motions. 

This raises the question of whether the velocity fields that we
observe in sharp-line A-type main sequence stars might also be due to
pulsations, but there are several reasons to doubt this. The first is
that the computations of Aerts et al. (\cite{Aertetal08}) do not show
any significant line {\em asymmetry}, the most persistent feature we
observed in the line profiles of A stars. Furthermore, we observe
depressed blue line wings and other symptoms of a velocity field in
stars significantly hotter than the hot limit of the instability
region (about 8700~K or a bit cooler; e.g. Joshi et
al. \cite{Joshetal03}). A third reason for rejecting pulsations is
that the asymmetry of line profiles in tepid stars has a constant
behaviour, reproducible through repeated observations of all the stars
in Table~\ref{detect-vels}: the excess absorption is always in the
short-wavelength line wing. It is difficult to see why pulsation would
not lead to excess absorption that would shift from one line wing to
the other. Finally, one of the stars in which the symptoms of a
non-thermal velocity field are strongest, HD~209625, has been observed
very carefully to search for evidence of pulsations (Carrier et
al. \cite{Carretal07}), but no hint of pulsational variations was
found. This is consistent with many observational studies which have
concluded that only marginal and evolved Am stars are found to pulsate
(cf.\ the review of Balona \cite{Balo04}), thus supporting theoretical
results from diffusion theory (Turcotte et al. \cite{Turcetal00}).  A
few exceptional classical Am stars, pulsating at very low amplitude,
have nevertheless been identified (Kurtz \cite{Kurt00}). In contrast,
not all the normal A stars in the centre of the instability strip are
found to pulsate above observational thresholds and $\delta$~Sct
variables with low \vsi\ are found to have a very large spread in
observed pulsation amplitudes (Breger \cite{Breg00}). Thus, although
we cannot completely exclude pulsation from being present in the stars
listed in Table~\ref{detect-vels} with $T_{\rm e} \leq 8700$~K, their
maximum variability level due to pulsation has to be very small,
making pulsation a very unlikely candidate for large velocity fields
that cause asymmetrically broadened line profiles.

It is of course possible to consider other explanations of the
depressed line wings than atmospheric motions. The excess broadening
is not caused by magnetic (Zeeman) broadening, as this would
selectively broaden lines of large Zeeman sensitivity; instead one
finds simply that all the strongest lines are broadened most. In
particular, if the broadening were magnetic, the strong
4634.07~\AA\ line of Cr {\sc ii} would be considerably narrower than
the nearby 4629.34~\AA\ line of Fe {\sc ii}; this is not
observed. Similarly, we can exclude the possibility that the effect is
caused by hyperfine splitting, because the principle isotopes of Fe,
Cr, and Ti, in which strong line distortion is observed, are
overwhelmingly even-even isotopes with no net nuclear magnetic moment,
and also because the observed effect depends strongly on stellar
\te. We can exclude the possibility that the selective broadening and
distortion is caused by rapid rotation seen pole on, as this would
broaden both strong and weak lines by a similar amount. We note that
also a direct interaction between the two members of each of the
observed binaries in the sample is unlikely to explain the asymmetric
profiles due to their constant behaviour over different rotational
periods and the varied nature of their binarity. Finally, it seems
unlikely that circumstellar material (which can have quite large
velocities) plays any role, as there is no observed emission at
H$\alpha$ and H$\beta$, such as one finds in Herbig AeBe stars.

We conclude that the atmospheric convective velocity field is
still the most probable explanation of the line profile peculiarities
reported here.

\section{Conclusions}

The results described in the preceding sections lead to a number of 
conclusions. 

Local atmospheric velocity fields are confidently detected in main
sequence stars having \te\ below 10000~K, but not in stars with higher
effective temperatures. These velocity fields are always detected via
clearly non-zero values of the microturbulence parameter $\xi$, which
is indistinguishable from zero for $\te \geq 10000$~K, but rises
quickly to a value near 2~\kms for lower \te. A larger value of $\xi$
is found essentially only for Am stars with \te\ between 7500 and
9000~K.

In all stars with sufficiently sharp lines (\vsi\ below about 6~\kms for
stars with $9\,000 \leq \te \leq 10\,000$~K, and below about 12 or
13~\kms\ for stars with $7\,500 \leq \te \leq 9\,000$~K), there is a
perfect correspondence between non-zero $\xi$ and other expressions of
a local velocity field. Sharp-line stars with non-zero microturbulence
also exhibit some or all of (1) best-fit values of \vsi\ that increase
with the equivalent width of the line, (2) significant line asymmetry,
with a deeper depression of the blue line wing for stars with
\te\ larger than about 7000~K, and a deeper red wing for cooler stars,
and (3) profiles of strong lines that are clearly not correctly
modelled by the usual LTE line formation theory if one assumes that
the only line broadening mechanisms are instrumental broadening,
thermal and pressure broadening, height-independent isotropic Gaussian
microturbulence, and rotation with \vsi\ derived from the weakest
available lines.

Our search for more stars which clearly exhibit characteristics (2)
and (3) has yielded only a few new examples (HD~2421A, HD~27962,
HD~103578A, HD~107168, HD~157486A, and probably HD~114330). However,
even this small number has tripled the available sample. The line
profile characteristics of these stars are consistent with those of
the very sharp-line stars already studied (HD~72660, HD~108642A, and
HD~209625; Landstreet \cite{jdl98}), both in form and in variation
with effective temperature. From the newly detected stars we learn
that the modifications to spectral lines produced by local velocity
fields are found in both Am and non-Am stars of similar \te, although
(consistent with the larger $\xi$ values found in Am stars) the
amplitude of line profile distortions seems larger in Am than in
normal A stars. The effects directly observable in really sharp-line
A-star spectra become stronger as \te\ decreases below 10\,000~K, down
to about 7\,000~K, below which the line profiles take on the already
well-known shape characteristic of solar-type stars. 

There are now sufficient data available to make useful comparison with
numerical simulations of convection for an interesting range of values
of \te\ and \logg, and the data for different tepid main sequence
stars show a reassuring degree of consistency. We should emphasize
that even if the effects we are studying here are observable only in
the most slowly rotating stars, they reveal that local line profiles
are not predicted correctly by standard theory, and aside from their
interest in allowing us to probe the atmospheric velocity fields
directly, these effects need to be understood (for example through
numerical simulations) so that they can be correctly incorporated into
spectrum synthesis codes for determining, for example, abundances in
stars of chemical elements for which only saturated spectral lines are
available.

It is observed (as is already well-known) that supergiants in this
\te\ range also show clear symptoms of local velocity fields. These
include large non-zero values of $\xi$, best fit \vsi\ values that
increase systematically with equivalent width, and profiles of strong
lines that are poorly fit by the canonical line model. However, such
stars do not appear to show the line {\em asymmetry} found in main
sequence A stars, and in the supergiants the line profile distortions
may be due to g-mode pulsations (Aerts et al. \cite{Aertetal08})
rather than to local convective velocities.

We conclude by summarising some of the principal observational
features that theoretical and numerical models should (and to some
extent do) account for.

One major regularity, and one of the strongest pieces of evidence
linking the observed line profile shapes to atmospheric convection, is
the variation of the strength of the effects discussed here with
effective temperature. As \te\ rises past about 7000~K, the line
profile forms observed change from the solar-type shapes to the shapes
observed here, with strongly depressed blue line wings. As
\te\ continues to rise, the effect becomes steadily weaker until at
about $\te \sim 10\,000$~K the line profiles become indistinguishable
from those predicted by the standard model of line shape. As discussed
by Landstreet (\cite{jdl98}), this behaviour is completely consistent
with the predictions of mixing-length theory. When mixing-length
theory is applied to model atmospheres appropriate to main sequence A
and late B stars, it is found that the strength of the super-adiabatic
temperature gradient in the atmosphere declines rapidly with
increasing \te, and its driving effect becomes negligible at about the
correct value of \te. Because this behaviour results from a
fundamental feature of A and B star atmospheres, namely the decline of
continuum opacity as H becomes more fully ionised with rising \te, it
is to be expected that it will also be found in numerical
models. However, because of the high computational expense of
modelling even a single \te\ value, this aspect of the velocity field
behaviour has not yet been studied numerically; most of the modelling
has been done for atmospheres with \te\ near 8000~K, where the effects
of the super-adiabatic temperature gradient in the lower line-forming
region are largest.

The form of profiles of strong lines, with excess absorption far in
the blue line wing relative to that in the red wing, represents the
most severe challenge to numerical modelling at present. The
observations strongly suggest that the integrated form of strong
absorption lines is produced by areas of gas rising with relatively
large velocity, while the descending gas is moving at a smaller
velocity, which presumably (to conserve mass) must cover a larger
fractional area. This is the opposite of the solar case, and is still
not found in numerical simulations in spite of serious efforts with 2D
and 3D codes (e.g. Steffen et al. \cite{sfl06}; Kochukhov et
al. \cite{Kochetal07}).

A further possible issue arises from the magnitude of the ``excess''
velocities detected, above line broadening due to thermal motions,
pressure broadening, ``standard'' microturbulence, macroscopic
broadening by rotation, and finite spectral resolution. These may be
as large as 10 -- 12~\kms\ as measured by comparing the horizontal
discrepancies between observed and computed line profiles. Since the
speed of sound is roughly 7.5 to 8.5~\kms\ through the line forming
region of a main sequence A star, the inferred local velocity field
may be significantly supersonic in places. This does not seem to be a
real problem, however; Freytag (\cite{Frey95}) and Kupka et
al.(\cite{Kupketal09}) both find that shocks occur in their
simulations, indicating clearly the local presence of supersonic
velocities.

It is very satisfying that line shape observations are able to provide
real tests and constraints that challenge efforts to model the
detailed behaviour of stellar convection. This confrontation will
certainly help to guide theory eventually to a more physically correct
description of the phenomenon in stars.

\begin{acknowledgements} 

We thank the referee for a number of useful suggestions and
comments. This work was supported by the Natural Sciences and
Engineering Research Council of Canada. Extensive use was made of the
Simbad database, operated at CDS, Strasbourg, France.

\end{acknowledgements}

\end{document}